\documentclass[fleqn,12pt,twoside]{article}
\newcommand{\bm}{\bibitem}

\def\be {\begin{equation}}
\def\ee {\end{equation}}
\def\bea {\begin{eqnarray}}
\def\eea {\end{eqnarray}}
\def\nn {\nonumber}

\usepackage[headings]{espcrc1}

\readRCS
$Id: espcrc1.tex,v 1.2 2004/02/24 11:22:11 spepping Exp $
\usepackage{graphicx}
\usepackage[figuresright]{rotating}

\newcommand{\AmS}{{\protect\the\textfont2
  A\kern-.1667em\lower.5ex\hbox{M}\kern-.125emS}}

\title{Collisional energy loss and the suppression of high $p_T$ hadrons
}

\author{Jan-e Alam\address{Variable Energy Cyclotron Centre, 
        1/AF Bidhan Nagar, Kolkata 700 064, India\\ 
        },%
        Abhee K. Dutt-Mazumder\address[SINP]{Saha Institute of Nuclear Physics, 
        1/AF Bidhan Nagar, Kolkata 700 064, India\\ 
}
                 and
       Pradip Roy\addressmark[SINP]}
       
\runtitle{collisional energy loss .. }
\runauthor{Alam et al.}

\begin{document}

\maketitle

\begin{abstract}
We calculate nuclear suppression factor ($R_{AA}$) for light hadrons
by taking only the elastic processes and argue that in the 
measured $p_T$ domain of RHIC, collisional rather than the 
radiative processes is the dominant mechanism for partonic energy loss.
\end{abstract}

\section{Introduction}

Jet quenching is one of the most promising tools to extract the initial
parton density produced in high energy heavy ion collisions~\cite{jetq}.
This is related to the final state energy loss of the leading partons 
causing depopulation of hadrons~\cite{npa757} 
at high transverse momentum ($p_T$).
The  suppressions of high $p_T$ hadrons and  unbalanced 
back-to-back azimuthal correlations of the dijet events
in Au+Au collisions measured at Relativistic Heavy Ion 
Collider (RHIC) provide experimental evidence in
support of the quenching. 
The observed nuclear suppression
of light hadrons ($\pi, \eta$) in $Au+Au$ collisions at
$\sqrt{s}=62-200$ A GeV at RHIC was reproduced by 
radiative loss only assuming that the contributions from
collisional loss is negligible.  However,
the non-photonic single electron spectrum from heavy meson decays measured by
PHENIX Collaboration~\cite{raacharm} put this assumption in question. 
No realistic parameter set can
explain this data using the radiative energy loss based jet tomography
model which either requires violation of bulk entropy bounds or 
non-perturbatively large $\alpha_s$ of the theory \cite{wick06}, or equivalently
one requires excessive transport co-efficient $\hat{q}_{\rm eff}= 14$ GeV$^2$/fm
\cite{armesto05}.  

In this light, we, in the present work would like to address if 
the omission of collisional loss at RHIC is justified or not. 
We argue that, whether the collisional or radiative loss is the main mechanism
of energy dissipation, is a $p_T$ dependent question. It also depends
on the energies of the colliding system and expected to be different
for RHIC and Large Hadron Collider (LHC). In contrast to the previous
works, we calculate nuclear suppression factor 
for pions at RHIC energy considering only the collision energy loss.

The partonic energy loss due to collisional
processes was revisited in~\cite{abhee05} and its importance
was demonstrated in the  context of RHIC  
in\cite{roy06}. It is shown 
in ref.\cite{abhee05} that there exists an energy range where collisional loss 
is as important as or even greater than its radiative counter part, hence
cannot be neglected in any realistic model of jet quenching. 
Recently this is also noted 
in ref.\cite{wick06,colloss}.

\section{Formalism}

We  use Fokker Planck (FP) equation
to dynamically evolve  parton spectra.
FP equation  can be derived from Boltzmann equation
if one of the partner of the binary collisions
is in thermal equilibrium and
the collisions are dominated by the small
angle scattering involving soft momentum exchange
\cite{roy06,alamprl94,svetitsky,FPE}.
For a longitudinally expanding plasma, 
FP equation reads:
\be
\left (\frac{\partial}{\partial t}
-\frac{p_\parallel}{t}\frac{\partial}{\partial p_\parallel}\right )f({\bf p},t)
=\frac{\partial}{\partial p_i}[p_i\eta f({\bf p},t)]\nn\\ 
+\frac{1}{2}
\frac{\partial^2}{\partial p_\parallel^2 }[{B_\parallel}({\bf p})f({\bf p},t)]
\nn\\
+\frac{1}{2}
\frac{\partial^2}{\partial p_\perp^2}[{B_\perp}f({\bf p},t)]
\label{fpexp}
\ee
where the second term on the left hand side arises due to 
expansion~\cite{baym}. 
Bjorken hydrodynamical model~\cite{bjorken1983}  
has been used here for space time evolution.
In Eq.~(\ref{fpexp}),  $f({\bf p},t)$ represents the non-equilibrium 
distribution of the partons under study, 
$\eta=(1/E)dE/dx$,
denotes drag coefficient,
$B_\parallel=d\langle(\Delta p_\parallel)^2\rangle/dt$,
$B_\perp=d\langle(\Delta p_\perp)^2\rangle/dt$, 
represent diffusion constants along parallel and
perpendicular directions of the propagating partons. 

In our calculations we include some of the important
features which were ignored in previous work~\cite{MT} in 
the context of jet quenching. These are: (i) the term 
corresponding to the longitudinal expansion is included,
(ii) the momentum evolution of parton distributions 
both along longitudinal and transverse direction is
considered and (iii) the mechanism of hadronization is
introduced.  

The transport coefficients, $\eta$, $B_\parallel$ and $B_\perp$
appeared in eq.(~\ref{fpexp}) have been calculated  in Ref.~\cite{roy06}.
The FP equation has been solved 
for the initial parton distributions 
parametrized as~\cite{mueller}:
\be
f(p_T,p_z,t=t_i)\equiv \frac{dN}{d^2p_Tdy}|_{y=0}=
\frac{N_0}{(1+\frac{p_T}{p_0})^\alpha},
\ee
where $p_0=1.75$, $\alpha=8$. 
Since asymptotically the multiplicity 
of any hadron species produced by a gluon jet is
$9/4$ times that of quark~\cite{ESW},   
the normalization
constant $N_0$ for gluons is chosen accordingly. 
Solving the FP equation with the boundary conditions, 
$f(\vec{p},t)\rightarrow 0$ for $|p|\rightarrow \infty$,
we evaluate 
the nuclear suppression factor, $R_{AA}$ defined as,
\bea
R_{AA}(p_T) 
= \frac{\sum_a \int f_a({\bf p^{\prime}}, \tau_c)|_{p_T^{\prime} = p_T/z}
D_{a/\pi^0}(z,Q^2)dz}{\sum_a \int f_a({\bf p^{\prime}},\tau_i)
|_{p_T^{\prime} = p_T/z}, 
D_{a/\pi^0}(z,Q^2)dz}
%\tau_i)D_{\pi^0/i}(z,Q^2)dz}
\eea
%%%%%%%%%%%%%%%%%%%%%%%%
\begin{figure}[htb]
\begin{minipage}[t]{80mm}
\includegraphics[scale=0.4]{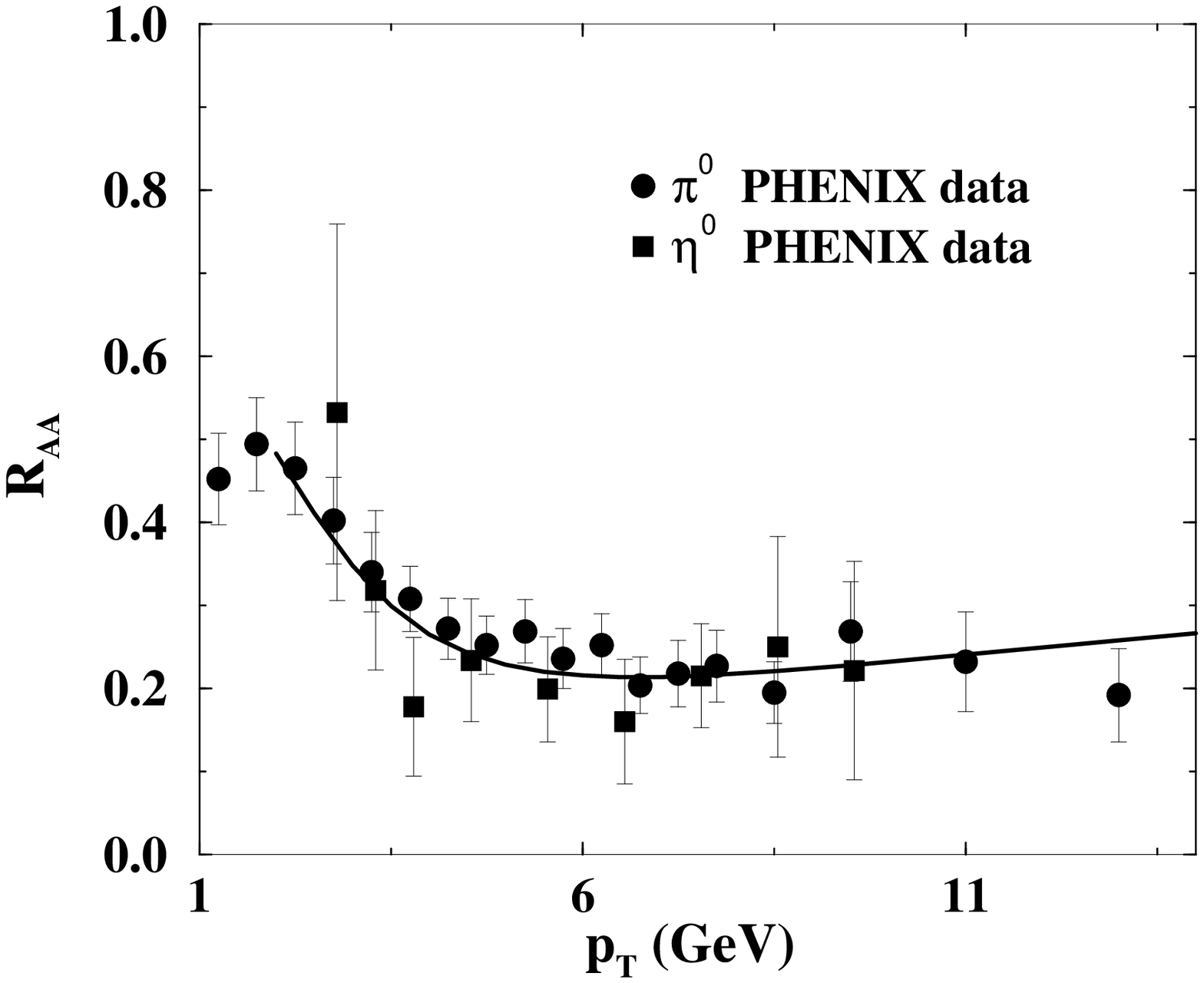}\\ 
%\framebox[74mm]{\rule[-26mm]{0mm}{52mm}}
\caption{
Nuclear suppression factor for pion. Experimental data are taken from
PHENIX collaboration 
\protect\cite{nuclex06} 
for Au + Au collisions
at $\sqrt{s}=200$ GeV. Solid  line indicates result from the
present calculation with collisional energy loss of the partons
propagating through the plasma before fragmenting into pions.
}
\label{fig1}
\end{minipage}
%\end{figure}
%
\hspace{\fill}
%
%\begin{figure}[htb]
\begin{minipage}[t]{80mm}
\includegraphics[scale=0.4]{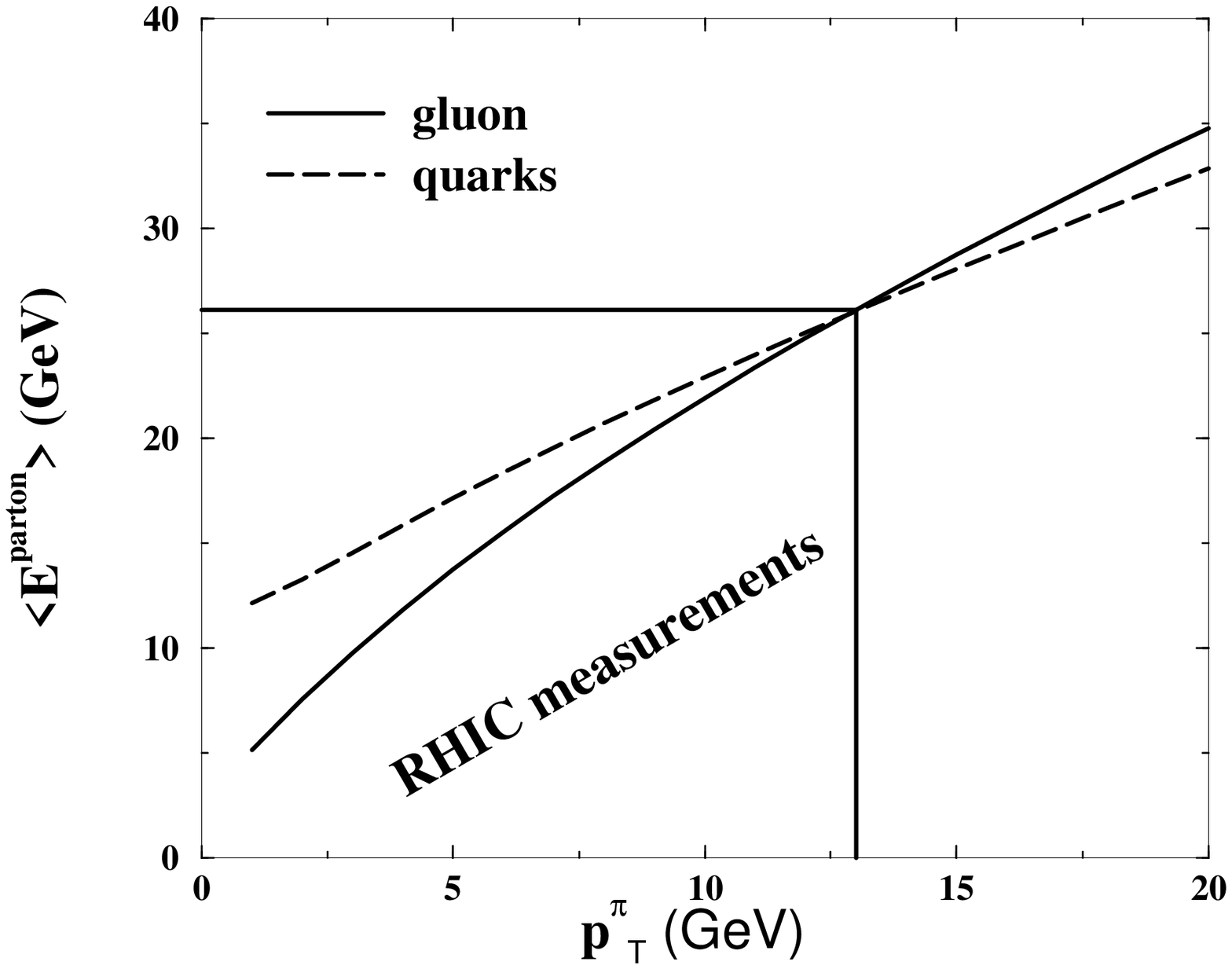}\\ 
%\framebox[74mm]{\rule[-26mm]{0mm}{52mm}}
\caption{Average parton energy versus transverse momentum of pion for 
$\sqrt{s}=200$ GeV/A.}
\label{fig2}
\end{minipage}
\end{figure}
%%%%%%%%%%%%%%%%%%%%%%%%
where $f({\bf p^\prime},\tau_i)$ and $f({\bf p^\prime},\tau_c)$ 
denote the   parton distributions at proper time $\tau_i$ and $\tau_c$ 
respectively. Here $\tau_i$ is the initial time and $\tau_c$ is the time
when the system cools down to the transition 
temperature $T_c$ (=190 MeV). 
We have taken $\alpha_s=0.3$ and the initial temperature $T_i=450$ MeV.

\section{Results}

The result of our calculations for neutral pion
is shown in 
Fig.~\ref{fig1} which describes the PHENIX data\cite{nuclex06} for $Au+Au$ at
$\sqrt{s}=200$ GeV reasonably well. 
To stress our point
further we also analyse the excitation function of the
nuclear suppression factor. This has recently been studied 
in~\cite{dEnterria05} attributing the suppression mechanism to 
radiative loss. As at lower
$\sqrt{s}$, collisional loss wins over its radiative counter part, 
we reanalyze the excitation function with the former and 
reproduce the data well~\cite{ard}.

To pin down the relative importance of 
$2 \leftrightarrow 2$ and $2\rightarrow 3$ processes, 
we  determine the average
energy of the parton which contribute to the measured 
$p_T$ window of the hadrons. 
The average fractional momentum,
$\langle z\rangle$ of 
the fragmenting partons carried by the pion is
calculated using relevant parton distribution and fragmentation functions,
{\i.e.} $\langle z\rangle\propto \int dz z \lbrace f(x_a,Q^2)f(x_b,Q^2) D_{a\rightarrow h}(z,Q^2)+
  f(x_b,Q^2)f(x_a,Q^2) D_{b\rightarrow h}(z,Q^2)\rbrace $.
For the parton distributions, we use CTEQ~\cite{cteq} 
including shadowing {\em via} EKS98 
parametrization\cite{eks}, while for the fragmentation function KKP 
parametrization is used \cite{kkp}.
The average energy of the parton, 
$\langle E^{\mathrm{parton}}\rangle$ is obtained by using the relation 
$<E^{parton}>=p_T^{\pi}/<z>$ for $y_{\pi} = 0$.
The results are shown in Fig.~\ref{fig2}.

It might be recalled that at RHIC energies the nuclear modification 
factor $R_{AA}(p_T)$ has been measured in the pion transverse 
momenta range $p_T\sim 1-13$ GeV. 
Assuming that these pions are originated from the fragmenting 
partons it is 
clear that the maximum average parton energy required is about 26 GeV here. 
We might compare this value with the determined $E_c$\footnote{Note
that  $E_c$ is defined to be the energy below which 
elastic loss dominates~\cite{abhee05}.} given in 
ref.~\cite{abhee05}
 and note that at these energies
collisional loss cannot be neglected. For lower beam energy,  62.4 (130) 
A GeV the value of maximum average parton energy required to produce
a 13 GeV pion is 16 (22) GeV, where 
the collisional loss will definitely be more important. 

\section{Summary}

In conclusion,
our investigations clearly suggest that in the measured
$p_T$ range of light hadrons at RHIC
collisional, rather than the radiative, is the dominant mechanism of 
jet quenching.  Inclusion of three body elastic channels 
might even increase $E_c$ making our point stronger. 

%{\bf Acknowledgment} We are grateful to David d'Enterria for providing us 
%the experimental data shown in  Fig. \ref{fig2}. 

\end{document}